\documentclass[aps,prl,twocolumn,showpacs]{revtex4}

\usepackage{graphicx}
\usepackage{mathtools}
\usepackage{empheq}
\usepackage{xcolor}

\begin{document}

\title{Inherent variability in the kinetics of autocatalytic protein self-assembly}

\author{Juraj Szavits-Nossan}
\email{jszavits@staffmail.ed.ac.uk}

\author{Kym Eden}
\email{k.eden@ed.ac.uk}

\author{Ryan J. Morris}

\author{Cait E. MacPhee}

\author{Martin R. Evans}

\author{Rosalind J. Allen}
\email{rallen2@staffmail.ed.ac.uk}
\affiliation{SUPA, School of Physics and Astronomy, University of Edinburgh, Mayfield Road, Edinburgh EH9 3JZ, United Kingdom}

\begin{abstract}
In small volumes, the kinetics of filamentous protein self-assembly is expected to show significant variability, arising from intrinsic molecular noise. This is not accounted for in existing deterministic  models. We introduce a simple stochastic model including nucleation and autocatalytic growth via elongation and fragmentation, which allows us to predict the effects of molecular noise on the kinetics of autocatalytic self-assembly. We derive an analytic expression for the lag-time distribution, which agrees well with experimental results for the fibrillation of bovine insulin. Our expression decomposes the lag time variability into contributions from primary nucleation and autocatalytic growth and reveals how each of these scales with the key kinetic parameters. Our analysis shows that significant lag-time variability can arise from both primary nucleation and from autocatalytic growth, and should provide a way to extract mechanistic information on early-stage aggregation from small-volume experiments.
\end{abstract}

\maketitle

% INTRODUCTION
%------------------------------------------------

The self-assembly of protein molecules into amyloid fibrils is associated with many degenerative diseases \cite{ChitiDobson06}, but also presents potential opportunities for the development of new materials \cite{GillamMacPhee13}. In both cases it is of outstanding importance to identify the specific microscopic steps responsible for amyloid aggregation, especially in its early stages. An important success of recent biophysical work has been to show that \textit{in vitro} kinetic data for amyloid fibril self-assembly can often be described by deterministic mechanistic models \cite{Wegner82,WegnerSavko82,Ferrone99,TanakaWeissman06,KnowlesDobson09,FerroneEaton80,BishopFerrone84,RuschakMiranker07,JeanVaux10}. However, it is unclear how far the results of these large-volume experiments can be translated to clinically relevant intracellular aggregation phenomena, which occur in far smaller volumes. 

\begin{figure}[!htb]
\includegraphics[width=8cm]{fig1ab.eps}
\includegraphics[width=8cm]{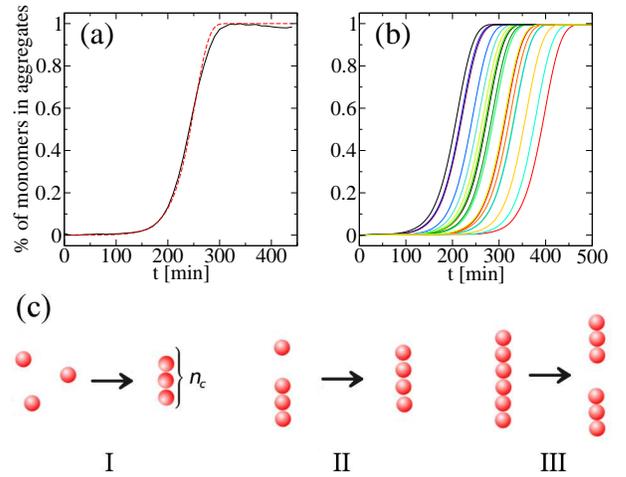}
\caption{\label{fig1} (a) Experimental kinetic curve (black line) for the aggregation of bovine insulin in a volume of $100$ $\mu$l from our own experiment, fitted to the theoretical prediction of a model \cite{KnowlesDobson09} involving primary nucleation, elongation and fragmentation (dashed red line); for full experimental details, see the Supplemental Material. (b) Kinetic curves obtained from kinetic Monte Carlo simulations of a stochastic version of the same model \cite{KnowlesDobson09,MorrisMacPhee13} and the fit parameters extracted from (a), but for a much smaller volume of $830$ fl. (c) Schematic illustration of: (I) primary nucleation, (II) elongation via polymerization and (III) fragmentation. The critical nucleus size for primary nucleation is denoted by $n_c$. }
\end{figure}

In large-volume {\em{in vitro}} experiments (typically 100-1000 $\mathrm{\mu}$l), measurements of the total mass of aggregated (fibrillar) protein as a function of time typically produce sigmoidal curves, as in figure \ref{fig1}(a) \cite{GillamMacPhee13}. These data show an initial lag phase in which no aggregated protein is detectable, followed by a rapid growth phase, terminating in a plateau once all the protein is in the aggregated form. In large volumes, these characteristic sigmoidal growth curves  can often be well fitted by deterministic kinetic models involving homogeneous primary nucleation (Fig. \ref{fig1}(c), I), filament elongation by monomer addition (Fig. \ref{fig1}(c), II) and autocatalysis via filament fragmentation (Fig. \ref{fig1}(c), III) \cite{KnowlesDobson09,Cohen11a,Cohen11b,Cohen11c,GillamMacPhee13} -- although  the contributions of primary nucleation and autocatalytic growth in the early stages of aggregation are often poorly distinguished \cite{BernackiMurphy09}. Importantly, these models lead to analytical predictions for scaling behavior; for example, if autocatalysis is dominant, the mean lag time scales as the inverse square root of the product of the protein concentration, elongation and fragmentation rates \cite{KnowlesDobson09}.

%----------
% Figure 1
%----------

In a clinical context, however, fibril formation happens in much smaller volumes, on the scale of a human cell (typically $500-3000$ fl). In small volumes, the stochastic nature of the underlying chemical reactions (``intrinsic molecular noise") is expected to become important, leading to inherent variability in the aggregation kinetics. Fig. \ref{fig1}(b) shows the results of replicate kinetic Monte Carlo simulations of a stochastic version of the autocatalytic growth model \cite{MorrisMacPhee13} in a volume of $830$ fl. These simulations predict significant variability in the lag time. Lag-time variability has also been observed in recent pioneering experiments on  bovine insulin fibril formation in micro-droplets  \cite{Knowles11}, as well as in classic small-volume experiments on the polymerization of sickle cell hemoglobin \cite{FerroneHofrichterEaton85a,FerroneHofrichterEaton85b,Hofrichter86}. Theoretical models which explain such data should provide a powerful tool for probing the mechanisms involved in early-stage aggregation. In particular, an important question concerns the relative roles of primary nucleation (Fig. \ref{fig1}(c), I) and autocatalytic growth (Fig. \ref{fig1}(c), II and III) in determining the lag-time distribution for amyloid fibril formation. So far, however, analytic predictions for lag-time distributions have been achieved only for models that do not fully take into account autocatalytic growth \cite{Szabo88,DLT12}.

In this paper, we present an analytic prediction for the lag-time distribution, for a stochastic model of filamentous protein self-assembly that includes primary nucleation, irreversible filament elongation and autocatalysis via fragmentation. We show that our prediction is in agreement with recent data for bovine insulin fibril formation in micro-droplets \cite{Knowles11}. This analytical solution allows a decomposition of the lag time variability into contributions from primary nucleation and autocatalytic growth, and reveals how each of these scales with the key kinetic parameters. 

%------------------------------------------------
% MODEL 
%------------------------------------------------

\textit{A Coarse-Grained Model for Autocatalytic Protein Self-Assembly.} Deterministic kinetic models for amyloid fibril self-assembly usually consist of dynamical equations for the mean number of fibrils $\langle n_i\rangle$ of a given length $i\geq n_c$, where $n_c$ is the size of the smallest growth-competent fibril (the  ``nucleus'') \cite{KnowlesDobson09,Cohen11a,Cohen11b,Cohen11c,GillamMacPhee13} \cite{note2}. For a model including homogeneous nucleation, irreversible elongation and fibril fragmentation (Fig. \ref{fig1}(c), I-III), these equations are non-linear, but various approximations have been successfully employed to obtain their full time-dependent solution \cite{KnowlesDobson09,Cohen11a,Cohen11b,Cohen11c}. The stochastic version of this model, where the number of each species $n_i$ is allowed to fluctuate is, however, analytically intractable, although it can be simulated numerically as we have done in Fig. \ref{fig1}(b). 

To obtain an analytic prediction for the lag-time distribution, we coarse-grain the model, while retaining the key processes of nucleation, elongation and fragmentation. Rather than tracking the full distribution of fibril lengths, we  track only the total number of fibrils $n=\sum_i n_i$ and the number of monomers in aggregates $m=\sum_i i n_i$, but treat them as discrete random variables, which can fluctuate due to intrinsic noise. This coarse-graining, which amounts essentially to summing over fibril lengths in the full model \cite{note3}, results in the following set of possible transitions between states $n,m$ of the system:
\begin{subequations}
\label{simple_model}
\begin{empheq}[left={n,m\rightarrow}\empheqlbrace]{align}
& n+1,m+n_c && \textrm{at rate $\alpha[c(t)]/\epsilon$}\label{nucleation_simple}\\ 
& n,m+1 && \textrm{at rate $2k_+ c(t) n$}\label{elongation_simple}\\
& n+1,m && \textrm{at rate $k_f m$}\label{fragmentation_simple}
\end{empheq}
\end{subequations} 

\noindent Primary nucleation is modeled by (\ref{nucleation_simple}) as an one-step process in which a new filament (called a ``nucleus'') is created instantaneously from $n_c$ free monomers at rate $\alpha[c(t)]\epsilon$. The rate $\alpha$ is assumed to depend on the molar concentration of free monomers $c(t)$ and $\epsilon=1/(V N_A)$, where $V$ is the volume and $N_A$ is Avogadro's constant. Transition (\ref{elongation_simple}) represents filament growth by monomer addition at rate $2k_+c(t)$; the factor of $2$ accounts for the fact that filaments can grow at both ends. Transition (\ref{fragmentation_simple}) represents fragmentation; this amounts to an autocatalytic creation of new fibrils from existing ones at rate $k_f$; the probability that any given fibril breaks is assumed to be proportional to its length. Although this assumption is somewhat simplistic \cite{Lee09}, we have also studied a model where fibrils break more frequently at their ends \cite{Paturej11}. This latter model, which is presented in the Supplemental Material, also obeys detailed balance by including backward reactions such as re-joining of fragmented fibrils and loss of monomers at fibril ends; however, none of these changes were found to affect the early-stage aggregation phenomena studied here \cite{SchreckYuan13,HongYong13}. In the rest of the paper we will further simplify the model by neglecting monomer depletion, which amounts to approximating the free monomer concentration $c(t)$ by $c_{\mathit{tot}}$; this has little effect on the lag phase.

 The probability distribution $P_{n,m}(t)$ for a given $n$ and $m$ obeys the following master equation
\begin{equation}
\label{master}
\begin{aligned}
\frac{d}{dt}P_{n,m}&=(\alpha/\epsilon) P_{n-1,m-n_c}+\mu n P_{n,m-1}\\
&+\lambda m P_{n-1,m}-(\alpha/\epsilon+\mu n+\lambda m)P_{n,m},
\end{aligned}
\end{equation}

\noindent where $\alpha\equiv\alpha(c_{\mathit{tot}})$, $\mu\equiv 2k_+c_{\mathit{tot}}$ and $\lambda\equiv k_f$. Starting with an initial condition $P_{n,m}(0)=\delta_{n,n_0}\delta_{m,m_0}$, we aim to solve for $P_{n,m}(t)$, and then to find the probability distribution for the lag time, i.e. for the time needed for the number of aggregated monomers $m$ to reach some predefined threshold $m_T$, which we define as $10\%$ of the total number of monomers (which is given by $c_{\mathit{tot}}/\epsilon$, assuming that $c_{\mathit{tot}}$ is measured in moles per unit volume).

%------------------------------------------------
% ANALYTIC SOLUTION IN THE LNA 
%------------------------------------------------

\textit{Analytic Solution for the Probability Distribution $P_{n,m}$.}  In order to obtain an analytic solution, we replace the master equation (\ref{master}) with a corresponding Fokker-Planck equation via the linear noise approximation (LNA), also known as Van Kampen's system size expansion \cite{NGvanKampen,Gardiner}. The LNA assumes that $n$ and $m$ can be decomposed into deterministic and fluctuating parts, 
\begin{subequations}
\begin{align}
n&=N_A V \phi(t)+\sqrt{N_A V}x_1\label{x_1}\\
m&=N_A V \psi(t)+\sqrt{N_A V}x_2\label{x_2},
\end{align}
\end{subequations}
\noindent where the fluctuating parts $x_1$ and $x_2$ are scaled by $\sqrt{N_A V}$, and are assumed to be small compared to the deterministic terms. The deterministic parts $\phi(t)$ and $\psi(t)$, expressed in units of concentration (here moles per unit volume), solve the following differential equations:
\begin{subequations}
\begin{align}
&\frac{d\phi}{dt}=\lambda\psi+\alpha,\quad \phi(0)=\epsilon n_0\equiv\phi_0\label{phi}\\
&\frac{d\psi}{dt}=\mu\phi+\alpha n_c,\quad \psi(0)=\epsilon m_0\equiv\psi_0.\label{psi}
\end{align}
\end{subequations}
\noindent Equations (\ref{phi}) and (\ref{psi}) may be solved to yield
\begin{subequations}
\begin{align}
&\phi(t)=\sqrt{\frac{\lambda}{\mu}}\Psi_0\mathrm{sinh}(\tau)+\Phi_0\mathrm{cosh}(\tau)-\frac{\alpha n_c}{\mu}\label{phi_sol}\\
&\psi(t)=\sqrt{\frac{\mu}{\lambda}}\Phi_0\mathrm{sinh}(\tau)+\Psi_0\mathrm{cosh}(\tau)-\frac{\alpha}{\lambda}.\label{psi_sol}
\end{align}
\end{subequations}
\noindent where we have adopted the following notation: $\tau=\sqrt{\mu\lambda} t$, $\Phi_0=\phi_0+\alpha n_c/\mu$ and $\Psi_0=\psi_0+\alpha/\lambda$. Equations (\ref{phi_sol}) and (\ref{psi_sol}) describe the time evolution of the mean concentrations of fibrils and aggregated protein, respectively, at early times. Solving $\psi(T)=m_T\epsilon$, where $m_T$ is the threshold concentration, yields the mean lag time $T$
\begin{equation}
T=\frac{1}{\sqrt{\mu\lambda}}\textrm{ln}\frac{D+\sqrt{D^2-\Psi_{0}^2+(\mu/\lambda)\Phi_{0}^2}}{\Psi_0+\sqrt{\mu/\lambda}\Phi_0},
\label{T}
\end{equation}
\noindent where $D=\alpha/\lambda+m_T\epsilon$. Eq. (\ref{T}) is a good approximation to the lag time reported in \cite{KnowlesDobson09} and predicts the same $T\propto (k_+ c_{\mathit{tot}} k_f)^{-1/2}$ scaling.

To determine the effects of intrinsic noise, we now turn to the fluctuating parts $x_1$ and $x_2$, which are governed by the following Fokker-Planck equation for the probability density $P(x_1,x_2,t)$,
\begin{equation}
\frac{\partial P}{\partial t}=-\sum_i\frac{\partial}{\partial x_i}(A_i P)+\frac{1}{2}\sum_{i,j}\frac{\partial^2}{\partial x_i x_j}(B_{ij}P)
\label{fp}
\end{equation}
\noindent where we assumed that $P(x_1,x_2,0)=\delta(x_1)\delta(x_2)$. The drift vector $\vec{A}$ and the diffusion matrix $B$ are given by
\begin{equation}
\vec{A}=\left( \begin{array}{c}\lambda x_2\\ \mu x_1\end{array}\right),\quad 
B=\begin{pmatrix}
\lambda\psi+\alpha & \alpha n_c\\
\alpha n_c & \mu\phi+\alpha n_{c}^{2} 
\end{pmatrix}.
\end{equation}

\noindent Equation (\ref{fp}) describes a two-variable  (time-dependent) Ornstein-Uhlenbeck process which can be solved 
by standard techniques \cite{Gardiner} and yields a bivariate Gaussian distribution with zero mean and time-dependent covariance matrix $\Sigma_{ij}=\langle x_i x_j\rangle$. To calculate the lag time distribution we only need to know $\Sigma_{22}=\langle x_2(t)^2\rangle$; the time-dependence of the other matrix elements can be found in the Supplemental Material.

%------------------------------------------------
% LAG TIME DISTRIBUTION 
%------------------------------------------------

\textit{Lag Time Distribution.}  
Building on these results, we now obtain an analytic expression for the lag time distribution $L(t)$. This is essentially a first-passage time problem; to  calculate $L(t)$, we look for all events such that $m$ has \textit{just} exceeded $m_T$ at a time $t$, given that it will exceed $m_T$ eventually,
\begin{equation}
\label{L}
L(t)=\frac{\frac{d}{dt}\text{Prob}[m>m_T,t]}{\text{Prob}[m>m_T,t\rightarrow\infty]}.
\end{equation}

\noindent The probability $\text{Prob}[m>m_T,t]$ can easily be calculated by integrating $P(x_1,x_2,t)$ and reads
\begin{equation}
\label{prob_mt}
\text{Prob}[m>m_T,t]=\frac{1}{2}\text{erfc}\left(\frac{m_T\epsilon-\psi(t)}{\sqrt{2\epsilon\langle x_2(t)^2\rangle}}\right),
\end{equation}
\noindent A lengthy but straightforward calculation for $\langle x_2(t)^2\rangle$ gives
\begin{equation}
\label{v2}
\begin{aligned}
&\langle x_2(t)^2\rangle=\textrm{cosh}(2\tau)\left[\frac{1}{6}\left(\Phi_0\frac{\mu}{\lambda}+\Psi_0\right)+\frac{\alpha n_c}{2\lambda}\right]\\
&+\textrm{sinh}(2\tau)\sqrt{\frac{\mu}{\lambda}}\left[\frac{\Phi_0+\Psi_0}{3}+\frac{\alpha n_c (n_c-1)}{4\mu}\right]\\
&+\frac{\textrm{cosh}\tau}{3\lambda}\left(\lambda\Psi_0-2\mu\Phi_0\right)+\frac{\textrm{sinh}\tau}{3}\sqrt{\frac{\mu}{\lambda}}\left(\Phi_0-2\Psi_0\right)\\
&-\frac{\alpha n_c (n_c-1)t}{2}+\frac{1}{2}\left(\Phi_0\sqrt{\frac{\mu}{\lambda}}-\Psi_0-\frac{\alpha n_c}{\lambda}\right).
\end{aligned}
\end{equation}

It now proves useful to introduce a new variable $r(t)$,
\begin{equation}
r(t)=\frac{\psi(t)-m_T\epsilon}{\sqrt{\epsilon\langle x_2(t)^2\rangle}}
\end{equation}
\noindent which measures the deviation of the mean fibril concentration ($\psi(t)$) from the threshold ($m_T\epsilon$), scaled by the root mean square of $m\epsilon-\psi(t)$. Using this variable, we combine expressions (\ref{L}), (\ref{prob_mt}) and (\ref{v2}) to  give our central result: an analytical expression for the lag time distribution in the linear noise approximation of the master equation (\ref{master}), which takes the form of a Gaussian in $r$ in the range $-\infty<r<r(\infty)$ \cite{note3},
\begin{equation}
\label{ltd}
L(t)dt=\frac{dr/dt}{\sqrt{2\pi}Z}e^{-\frac{r(t)^2}{2}}dt=\frac{1}{\sqrt{2\pi}Z}e^{-\frac{r^2}{2}}dr,
\end{equation}
\noindent where $Z=\textrm{erfc}(-r(\infty)/2)$. Importantly, Eq.~(\ref{ltd}) allows us to easily calculate moments of the lag time distribution. For example, to calculate the  mean lag time $\langle t\rangle$ and its standard deviation $\sigma$, we express $t$ and $t^2$ as functions of $r$ and perform a Taylor expansion around $r=0$ (see the Supplemental Material for details). This gives 
\begin{equation}
\langle t\rangle\approx T\quad\text{and}\quad\sigma\approx\frac{\sqrt{\epsilon\langle x_{2}^{2}(T)\rangle}}{\mu\phi(T)+\alpha n_c}.
\label{stdev_lt}
\end{equation}
For most proteins, the fragmentation rate $\lambda \equiv k_f$ is much smaller than the net fibril elongation rate $\mu \equiv 2k_+ c_{tot}$; i.e.  $\lambda\ll\mu$. If we also assume that no fibrils are present at time $t=0$ ($\phi_0=\psi_0=0$), we can write a simpler expression for the standard deviation of the lag time, 
\begin{equation}
\sigma=\frac{(2/3)^{1/2}}{(\mu\lambda)^{1/4}(\alpha N_A V)^{1/2}}.
\label{sigma_simple}
\end{equation}
Remarkably, Eq.~(\ref{sigma_simple}) implies that the lag time variance scales in a simple way with the model parameters. Like the mean lag time, the variance is predicted to scale as $\sqrt{\mu \lambda}\sim\sqrt{k_f k_+}$. Interestingly, however, the mean and variance of the lag time may show different dependencies on the protein concentration $c_{tot}$; while the mean scales as $c_{tot}^{-1/2}$, in expression (15) for the variance this factor (which arises from $\mu$) is multiplied by an additional factor due to the $c_{\mathit{tot}}$-dependent nucleation rate $\alpha$; the scaling of this factor depends on the nucleus size $n_c$.

\begin{figure}[!hbt]
\includegraphics[width=8cm]{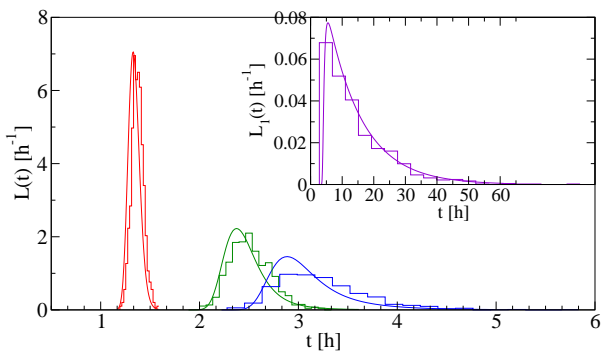}
\caption{\label{fig2} The lag-time distribution $L(t)$ for several values of $\alpha$, compared to that obtained by running $1000$ independent kinetic Monte Carlo simulations of the full stochastic model (in which individual fibril lengths are resolved) \cite{MorrisMacPhee13}. From left to right: $\alpha=50$ (full line), $5$ (dashed line), $1.5$ (dot-dashed line), all in units of $10^{-15}$ $\textrm{mol}/(\textrm{ls})$. Inset: $L_1(t)$ (dashed line) compared to simulations for $\alpha=5\cdot 10^{-17}$ $\textrm{mol}/(\textrm{ls})$. The other parameters are: $V=830$ fl, $M_T=10\%$ of $c_{\mathit{tot}}$, $c_{\mathit{tot}}=100$ $\mu$mol/l, $n_c=2$, $k_+=5\cdot 10^{4}$ l/(mol s) and $k_f=3\cdot 10^{-8}$ $\textrm{s}^{-1}$.}
\end{figure}

It is important to note that results (\ref{ltd})-(\ref{sigma_simple}) only hold in the regime dominated by growth, where fluctuations in $n$ and $m$ are much smaller than their averages, \emph{for all times}. In contrast, for slow nucleation rates, a significant portion of the lag time is spent waiting for the first nucleus to be spontaneously created, which is a fluctuation-driven process. We take this into account by convolving $L(t-t')$ with the waiting time distribution $(\alpha/\epsilon)\exp(-\alpha t'/\epsilon)$ for the primary nucleation event, to give
\begin{equation}
L_1(t)=(\alpha/\epsilon)\int_{0}^{t}dt' e^{-(\alpha/\epsilon) t'}L(t-t'),
\label{convolution}
\end{equation}
\noindent where in the expression for $L(t-t')$ we set $\phi(t')=\epsilon$ and $\psi(t')=n_c\epsilon$ (i.e. assume one fibril of size $n_c$ at time $t'$). Figure \ref{fig2} shows that the lag-time distributions predicted by Eqs.~(\ref{ltd}) and (\ref{convolution}) are in good agreement with the results of stochastic simulations of the full model (which takes into account fibril lengths), for several values of the primary nucleation rate $\alpha$. For relatively fast nucleation rates, our ``bare'' LNA prediction $L(t)$ (Eq. \ref{ltd}) is sufficient (main plots in Figure \ref{fig2}); for slower nucleation rates (inset in Fig. \ref{fig2})), (\ref{convolution}) should be used instead (inset to Fig. \ref{fig2}).

%----------
% Figure 2
%----------

For slow nucleation rates, we can separate the contributions of primary nucleation and autocatalytic growth to the lag time variance in a simple way. Assuming $L(t)$ can be replaced by a Gaussian in $t$, we can use (\ref{stdev_lt}) to compute the integral in (\ref{convolution}) in a closed form which reveals that $L_1(t)$ has mean $T_1$ and standard deviation $\sigma_1$ given by
\begin{equation}
T_1=\frac{\epsilon}{\alpha}+T,\quad \sigma_1=\sqrt{\left(\frac{\epsilon}{\alpha}\right)^{2}+\sigma^2}.
\label{sigma1_simple}
\end{equation}
Thus the lag-time variance is given by a simple sum of the variance of the exponential waiting time distribution for the primary nucleation event, and the contribution from autocatalytic growth, given by Eq.~(\ref{sigma_simple}).

%----------
% Figure 3
%----------

\begin{figure}[!htb]
\includegraphics[width=8cm]{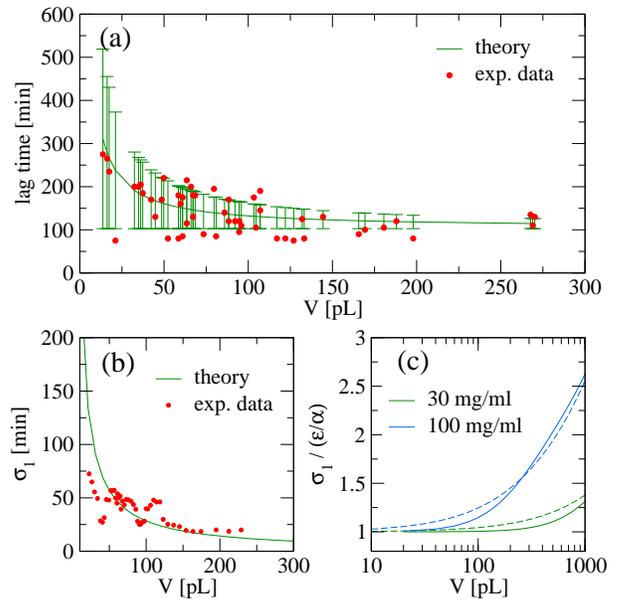}
\caption{\label{fig3} (a) Volume dependence of the lag time for the aggregation of bovine insulin in microdroplets of varying volume (red dots) \cite{Knowles11}, compared to the mean (solid line) and the standard deviation (error bars) from  Eq.~\ref{sigma1_simple}, using the following values, which were obtained from Ref. \cite{Knowles11}: $T=104$ min, $\alpha=1/(1.7\cdot 10^{-7}N_A)$ mol/(l s), $k_+=8.9\cdot 10^{4}$ l/(mol s), $k_f=2\cdot 10^{-8}$ $\textrm{s}^{-1}$ and assuming $n_c=2$. (b) Corresponding volume dependence of the standard deviation; the red dots are 6-point moving standard deviation from the experimental data, and the solid green line is from (\ref{sigma1_simple}). (c) Theoretical predictions for the standard deviation $\sigma_1$ as a function of volume, relative to $\epsilon/\alpha$. The green (lower) lines correspond to the protein concentration $30$ mg/ml used in Ref. \cite{Knowles11}, while the blue (upper) lines are for a higher protein concentration, $100$ mg/ml, assuming that $\alpha(c_{\mathit{tot}})\propto{c_\mathit{tot}}^{n_c}$ \cite{KnowlesDobson09}. In both cases, the dashed lines correspond to (\ref{sigma1_simple}) while the solid lines are calculated numerically from (\ref{convolution}).}
\end{figure} 

%------------------------------------------------
% EXPERIMENTS
%------------------------------------------------

\textit{Comparison with Experimental Results for Bovine Insulin.} So far, the only available experimental data on amyloid fibril nucleation in small volumes is that of Knowles {\em{et al.}}, who tracked the fibrillation of bovine insulin in 52 micro-droplets of volumes in the range $10-300$ pl, using ThT fluorescence \cite{Knowles11}. Fig. \ref{fig3}(a) shows the resulting lag times (red dots) as a function of droplet volume, compared to our theoretical prediction; the green line shows the mean lag time $T_1$, and the error bars show the standard deviation $\sigma_1$, from Eq.~(\ref{sigma1_simple}). No fitting parameters were used in this plot; rather the parameters $k_+$, $k_f$ and $\alpha$ were taken directly from the measurements of Ref. \cite{Knowles11} \cite{note5}. While there are not enough experimental data points to plot a lag-time distribution for any given volume, Fig. \ref{fig3}(a) shows that the variability observed in the experiments is consistent with our theory. This is further evidenced in Fig. \ref{fig3}(b), where we plot directly the volume-dependence of the standard deviation. 

We can also use our result, Eq.~(\ref{sigma1_simple}), to explore the relative contributions of primary nucleation and autocatalytic growth to the lag-time variability. Fig. \ref{fig3}(c) shows our theoretical prediction for the standard deviation $\sigma_1$, relative to that for primary nucleation only, $\epsilon/\alpha$. The relative contribution of autocatalytic growth increases strongly as the volume increases (although the total variability decreases with $V$). For the protein concentration of $30$ mg/ml used in Ref. \cite{Knowles11}, primary nucleation is the main contributor. However, for higher protein concentrations, we predict that autocatalytic variability becomes significant even at smaller volumes, on the scale of a human cell.

%--------------------------------------------------------------------------
% CONCLUSIONS
%--------------------------------------------------------------------------

\textit{Conclusion.} We have presented an analytic expression for the lag time distribution, for a stochastic model of autocatalytic protein self-assembly which includes nucleation, elongation and fragmentation. Our solution provides simple scaling relations for the contributions to lag-time variability due to primary nucleation and autocatalysis, both of which can be significant under realistic conditions. The implications of molecular noise for variability in clinical outcomes between individuals, as well as the possible connection to variability between replicates in large volume experiments \cite{XueRadford08} present interesting and important directions for future work.

%--------------------------------------------------------------------------
% ACKNOWLEDGMENTS
%--------------------------------------------------------------------------
\begin{acknowledgments}
We thank Tuomas Knowles for kindly providing the data from Ref. \cite{Knowles11} which we have used in Fig. \ref{fig3} and Line Jourdain for discussions in the early stages of this work.  KE and JSN contributed equally to this work. This research was supported by EPSRC under grant number EP/J007404/1. KE was supported by an EPSRC DTA studentship and RJA was supported by a Royal Society University Research Fellowship.
\end{acknowledgments}

%--------------------------------------------------------------------------
% REFERENCES
%--------------------------------------------------------------------------

%--------------------------------------------------------------------------
% SUPPLEMENTAL MATERIAL
%--------------------------------------------------------------------------

\onecolumngrid
\newpage

%\maketitle
\begin{center}
  {\Large \bf  Supplemental Material to:\\ Inherent variability in the kinetics of autocatalytic protein self-assembly}
\end{center}
\medskip

\allowdisplaybreaks[1]

\setcounter{equation}{0}
\setcounter{figure}{0}
\renewcommand{\thefigure}{S\arabic{figure}}
\renewcommand{\theequation}{S\arabic{equation}}

%------------------------------------------------
% EXPERIMENTS
%------------------------------------------------

\section{Experimental setup}

To obtain the experimental data shown in Fig. 1(a), bovine insulin was obtained from Sigma-Aldrich (I$55004$, lot number $0001434060$). The zinc content was approximately 0.5\% (w/w). The samples in this study were dissolved in $25$ mM HCl (pH $1.6$) immediately prior to the experiment. All solvents and solutions were filtered through a 0.22 $\mu$m filter (Millipore). Concentrations were checked via UV-Vis absorption spectroscopy. ThT was added to each solution to a final concentration of $20$ $\mu$M. Experiments were conducted using Corning NBS $96$-well plates (Corning $3641$). These plates are coated with a proprietary PEO-like hydrophilic coating which minimizes the interaction of protein with the plates. It was found that using these plates significantly minimized the variability in the kinetics compared to standard polystyrene microwell plates. Each well of the plate was filled with 100 $\mu$L of solution.  Experiments were replicated across $2$-$3$ whole plates for each protein concentration. The plates were sealed with a plastic adhesive and then incubated at 60$^{\circ}$ C. The kinetics of aggregation was followed via the binding of the fluorescent dye Thioflavin T (ThT), which binds preferentially to the fibrillar form of the protein. Fluorescence readings were taken from the bottom optic. The final number of individual experiments for a given protein concentration ranged from $\sim$ 140-200.

In total, $68$ replicate kinetic curves at protein concentration of $0.75$ mg/mL, at pH $1.6$ and 60$^{\circ}$ C were used to obtain the average growth curve in Fig. 1(a) (solid black line), which was then fitted to the theoretical prediction from \cite{KnowlesDobson09} (dashed red line)
\begin{equation}
\label{fibrilmass_knowles}
\frac{M(t)}{c_{\mathit{tot}}}=1-\exp\left\{-\frac{\alpha}{c_{\mathit{tot}}}\left[\textrm{cosh}(\sqrt{2k_+c_{\mathit{tot}}k_f})-1\right]\right\},
\end{equation}

\noindent to obtain the kinetic parameters $n_c=2$, $\alpha=5.8\cdot 10^{-16}$ mol/(l s), $k_+=5\cdot 10^{4}$ l/(mol s) and $k_f=3\cdot 10^{-8}$ $\textrm{s}^{-1}$. These parameters were used in the simulations of Fig. 1(b).

%------------------------------------------------
% FULL STOCHASTIC MODEL INVOLVING NUCLEATION, 
% ELONGATION AND FRAGMENTATION
%------------------------------------------------

\section{Full stochastic model involving nucleation, elongation and fragmentation}

The full stochastic model \cite{MorrisMacPhee13} tracks the number of fibrils $n_i$ for each fibril length $i\geq n_c$, where $n_c$ is the size of the smallest stable fibril. The state of the system is then fully described by the collection of integers $\{n_{n_c},n_{n_c+1}\dots,\}$. 

The master equation that governs the time evolution for the probability $P(\{n_i\},t)$ to find the system in a state $\{n_i\}$ is then given by
\begin{equation}
\label{master_full}
\begin{aligned}
\frac{d}{dt}P(\{n_i\})&=\frac{\alpha(c(t)+n_c\epsilon)}{\epsilon}\theta(n_{n_c}-1)P(\{n_{n_c}-1,\dots\})-\frac{\alpha(c(t))}{\epsilon}P(\{n_i\})\\
&+2k_+(c(t)+\epsilon)\sum_{i\geq n_c}(n_i+1)P(\dots,n_i+1,n_{i+1}-1,\dots)\theta(n_i+1)\theta(n_{i+1}-1)\\
&-2k_+c(t)\sum_{i\geq n_c}n_iP(\{n_i\})+\textrm{fragmentation terms}
\end{aligned}
\end{equation}

\noindent Here, $c(t)$ is the molar concentration of free monomers, $c(t)=c_{\mathit{tot}}-\sum_{i\geq n_c}i n_i\epsilon$, where $c_{\mathit{tot}}$ is the initial monomer concentration, $\epsilon=1/(V N_A)$, $V$ is the volume and $N_A$ is Avogadro's constant; to account for the allowed transitions between the states, we used the Heaviside step function $\theta(n)$ which equals $0$ for $n<0$ and $1$ for $n\geq 0$. The first two terms in Eq. (\ref{master_full}) describe nucleation, the next two elongation and the rest of terms, which we will write below, describe fragmentation. 

Now, let us define the total number of fibrils $n$ and the number of monomers in aggregates $m$, respectively,
\begin{equation*}
n=\sum_{i\geq n_c}n_i,\quad m=\sum_{i\geq n_c}i n_i.
\end{equation*}

\noindent The probability $P(n,m,t)$ to find the system with a particular $n$ and $m$ at time $t$ can be obtained by summing $P(\{n_i\})$ over all states $\{n_i\}$ having $\sum_{i\geq n_c}=n$ and $\sum_{i\geq n_c}i n_i=m$, i.e.
\begin{equation}
\label{prob_coarse_grained}
P(n,m,t)=\sum_{\{n_i\}}P(\{n_i\})\delta(n-\sum_{i\geq n_c}n_i)\delta(m-\sum_{i\geq n_c}i n_i),
\end{equation}

\noindent where $\delta(i,j)$ denotes the Kronecker delta function. 

In Eq. (\ref{master_full}), the nucleation and elongation terms both contain the number of free monomers $c(t)$, which is state-dependent. As we are interested in early times only, we can ignore monomer depletion and approximate $c(t)$ with $c_{\mathit{tot}}$. From there it is straightforward to obtain the master equation for $P(n,m,t)$, which reads
\begin{equation}
\label{master_coarse_grained}
\begin{aligned}
\frac{d}{dt}P(n,m,t)&=(\alpha/\epsilon)P(n-1,m-n_c,t)-(\alpha/\epsilon)P(n,m,t)+2k_+c_{\mathit{tot}}nP(n,m-1)\\
&-2k_+c_{\mathit{tot}}nP(n,m)+\textrm{fragmentation terms},
\end{aligned}
\end{equation}

\noindent where we have used the notation $\alpha\equiv\alpha(c_\mathit{tot})$.

To write the fragmentation terms, we have to distinguish between fibrils of length $n_c \leq i\leq 2 n_c-1$ and $i\geq 2n_c$. These two cases differ in the possible ways a fibril can be broken, taking into account the fact that the smallest stable fibril unit has length $n_c$. If a fibril of length $n_c\leq i<2n_c$ breaks into two fibrils, then at least one of them must be \emph{unstable}, and therefore will dissolve. On the other hand, if a fibril of length $i\geq 2n_c$ breaks into two fibrils, at least one of them must be \emph{stable}. That said, we can write the fragmentation terms on the r.h.s. of the master equation (\ref{master_full}) as
\begin{equation}
\label{fragmentation}
\begin{aligned}
\textrm{fragmentation terms}&=\sum_{i=n_c}^{2n_c-1}\sum_{k=1}^{i-n_c}k_f(n_i+1)P(\dots,n_{i-k}-1,\dots,n_i+1,\dots)\theta(n_{i-k}-1)\\
&+\sum_{i=n_c}^{2n_c-1}\sum_{k=n_c}^{i-1}k_f(n_i+1)P(\dots,n_{k}-1,\dots,n_i+1,\dots)\theta(n_{k}-1)\\
&+\sum_{i\geq 2n_c}\sum_{k=1}^{n_c-1}k_f(n_i+1)P(\dots,n_{i-k}+1,\dots,n_i+1,\dots)\theta(n_{i-k}-1)\\
&+\sum_{i\geq 2n_c}\sum_{k=i-n_c+1}^{i-1}k_f(n_i+1)P(\dots,n_{i-k}+1,\dots,n_i+1,\dots)\theta(n_{k}-1)\\
&+\sum_{i=n_c}^{2n_c-2}\sum_{k=i-n_c+1}^{n_c-1}k_f(n_i+1)P(\dots,n_i+1,\dots)\\
&+\sum_{i\geq 2n_c}\sum_{k=n_c}^{i-n_c}k_f(n_i+1)P(\dots,n_{k}-1,\dots,n_{i-k}-1,\dots,n_i+1,\dots)\theta(n_{k}-1)\theta(n_{i-k}-1)\\
&-\sum_{i\geq n_c}\sum_{k=1}^{i}k_fn_iP(\{n_i\})\theta(n_i-1)
\end{aligned}
\end{equation}

\noindent The first four terms above describe fragmentation events which produce exactly one unstable fibril; the fifth and the sixth term describe events that produce two unstable and two stable fibrils, respectively. Equation (\ref{master_full}) with the fragmentation terms given in (\ref{fragmentation}) is the basis of our full stochastic model, which is then compared to the analytical predictions of the simpler, coarse-grained model (in Fig. 2 of the main text).

\section{Coarse-graining the full stochastic model}

To obtain our coarse-grained model, we ignore the occurrences of unstable fibrils. This yields the following master equation for $P_{n,m}(t)$ after summing (\ref{master_full}) and (\ref{fragmentation}) over all states $\{n_i\}$ with given $n$ and $m$,
\begin{equation}
\label{master_coarse_grained_full}
\begin{aligned}
\frac{d}{dt}P(n,m,t)&=(\alpha/\epsilon)P(n-1,m-n_c,t)-(\alpha/\epsilon)P(n,m,t)+2k_+m_{\mathit{tot}}nP(n,m-1)\\
&-2k_+c_{\mathit{tot}}nP(n,m)+k_f[m-(2n_c-1)(n-1)]P(n-1,m)-k_f[m-(2n_c-1)n]P(n,m,t).
\end{aligned}
\end{equation}

\noindent Except for very early times, $m$ is expected to be much larger than $n$, and so $m-(2n_c-1)n$ can be approximated by $m$. This leads to the following master equation,
\begin{equation}
\label{master_coarse_grained_final}
\begin{aligned}
\frac{d}{dt}P(n,m,t)&=(\alpha/\epsilon)P(n-1,m-n_c,t)-(\alpha/\epsilon)P(n,m,t)+2k_+c_{\mathit{tot}}nP(n,m-1)\\
&-2k_+c_{\mathit{tot}}nP(n,m)+k_fmP(n-1,m)-k_fmP(n,m,t),
\end{aligned}
\end{equation}

\noindent which is the subject of our theoretical analysis.

%------------------------------------------------
% FULL STOCHASTIC MODEL RESPECTING DETAILED
% BALANCE
%------------------------------------------------

\section{Full stochastic model respecting detailed balance}

The coarse-grained model discussed in the text main text takes account of the following three chemical reactions:
\begin{eqnarray}
&& n_cm\xrightarrow[]{k_n}F_{n_c},\label{cr_nucleation}\\
&& m+F_i\xrightarrow[]{k_+}F_{i+1},\quad i\geq n_c,\label{cr_elongation}\\
&& F_{i+j}\xrightarrow[]{k_f} F_i+F_j,\quad i,j\geq n_c,\label{cr_fragmentation}
\end{eqnarray}
\noindent where $m$ denotes a monomer and $F_i$ a fibril of length $i$. This reaction scheme does not obey detailed balance. To extend our scheme so that it does obey detailed balance, we add the following 'backward' processes
\begin{eqnarray}
&& F_{n_c}\xrightarrow[]{k_d}n_c m,\label{cr_disintegration}\\
&& F_{i+1}\xrightarrow[]{k_-}m+F_{i},\quad i\geq n_c,\label{cr_chipping}\\
&& F_{i}+F_{j}\xrightarrow[]{k_c}F_{i+j},\quad i,j\geq n_c,\label{cr_coalescence}
\end{eqnarray}
\noindent which represent disintegration, depolymerization and fibril coalescence (end-joining), respectively. Here we assume length-independent kernels for the fragmentation and coalescence processes, which is true for rate-limited reactions. However, reactions (\ref{cr_fragmentation}) and (\ref{cr_chipping}) together can be understood as an example of inhomogeneous fragmentation in which a fibril is more likely to break at its ends ($k_->k_f$). In particular, we set $k_-=\kappa k_f$, where $\kappa\approx 1.1$ is chosen in accordance with a recent theoretical study on rupture probabilities along a single polymer chain \cite{Paturej11}. 

The rate equations for the processes (\ref{cr_nucleation})-(\ref{cr_coalescence}) read
\begin{eqnarray}
&&\frac{d}{dt}\rho_{1}=(n_c k_d \rho_{n_c}-n_c k_n \rho_{1}^{n_c})+\sum_{j=n_c}^{\infty}\left(2k_-\rho_{j+1}-2k_+\rho_1\rho_j\right),\label{n1}\\
&&\frac{d}{dt}\rho_{n_c}=(k_n \rho_{1}^{n_c}-k_d \rho_{n_c})+(2k_-\rho_{n_c+1}-2k_+\rho_1 \rho_{n_c})\nonumber\\
&& \qquad +\sum_{j=n_c}^{\infty}\left(2k_f \rho_{n_c+j}-2k_c \rho_{n_c}\rho_j\right),\label{nnc}\\
&& \frac{d}{dt}\rho_i=(2k_+\rho_1 \rho_{i-1}-2k_-\rho_i)+(2k_-\rho_{i+1}-2k_+\rho_1 \rho_i)\nonumber\\
&& \qquad +\sum_{j=n_c}^{\infty}\left(2k_f \rho_{i+j}-2k_c \rho_i \rho_j\right)+\sum_{j=n_c}^{i-n_c}\left(k_c \rho_i \rho_{i-j}-k_f \rho_i \right),\quad i\geq n_c+1\label{ni}
\end{eqnarray}
\noindent where $\rho_1$ denotes the monomer concentration and $\rho_i$ denotes the concentration of fibrils of length $i$ (all in units of molar). 

Detailed balance can then be implemented by demanding that each of the terms in parentheses vanishes for the equilibrium concentrations $\rho_{1}^{\textrm{eq}}$ and $\rho_{i}^{\textrm{eq}}$ for $i\geq n_c$:
\begin{eqnarray}
&& k_d \rho_{n_c}^{\textrm{eq}}=k_n (\rho_{1}^{\textrm{eq}})^{n_c}\label{detbal1}\\
&& k_-\rho_{i+1}^{\textrm{eq}}=k_+\rho_{1}^{\textrm{eq}}\rho_{i}^{\textrm{eq}},\quad i\geq n_c\label{detbal2}\\
&& k_f \rho_{i+j}^{\textrm{eq}}=k_c\rho_{i}^{\textrm{eq}}\rho_{j}^{\textrm{eq}},\quad i,j,\geq n_c\label{detbal3}.
\end{eqnarray}
\noindent These equations have the following steady-state solution for $i\geq n_c$
\begin{equation}
\rho_{i}^{\textrm{eq}}=\frac{k_n (\rho_{1}^{\textrm{eq}})^{n_c}}{k_d}\left(\frac{k_+\rho_{1}^{\textrm{eq}}}{k_-}\right)^{i-n_c},\quad i\geq n_c
\end{equation}
\noindent provided that
\begin{equation}
k_c=k_f\frac{k_d}{k_n}\left(\frac{k_+}{k_-}\right)^{n_c}.
\label{kc}
\end{equation}
\noindent From conservation of mass, $\rho_{1}^{\textrm{eq}}+\sum_{i=n_c}^{\infty}i\rho_{i}^{\textrm{eq}}=c_{\mathit{tot}}$, we also have that
\begin{equation}
\left[\frac{k_n (\rho_{1}^{\textrm{eq}})^{n_c}}{k_d}\right]\left[\frac{n_c}{1-k_+\rho_{1}^{\textrm{eq}}/k_-}+\frac{k_+\rho_{1}^{\textrm{eq}}/k_-}{(1-k_+\rho_{1}^{\textrm{eq}}/k_-)^2}\right]=c_{\mathit{tot}}-\rho_1,
\label{totalmass}
\end{equation}
\noindent which can be used to determine $k_d$, yielding 
\begin{equation}
k_d=\left[\frac{k_n (\rho_{1}^{\textrm{eq}})^{n_c}}{c_{\mathit{tot}}-\rho_1}\right]\left[\frac{n_c}{1-k_+\rho_{1}^{\textrm{eq}}/k_-}+\frac{k_+\rho_{1}^{\textrm{eq}}/k_-}{(1-k_+\rho_{1}^{\textrm{eq}}/k_-)^2}\right].
\label{kd}
\end{equation}
\noindent The only remaining unknown parameter is $\rho_{1}^{\textrm{eq}}$, which can be related to $\bar{l}$, the mean fibril length in equilibrium, by
\begin{equation}
\bar{l}=\frac{\sum_{i=n_c}^{\infty}i\rho_{i}^{\textrm{eq}}}{\sum_{i=n_c}^{\infty}\rho_{i}^{\textrm{eq}}}=n_c+\frac{k_+\rho_{1}^{\textrm{eq}}/k_-}{1-k_+\rho_{1}^{\textrm{eq}}/k_-},
\end{equation}
\noindent yielding
\begin{equation}
\rho_{1}^{\textrm{eq}}=\frac{k_-}{k_+}\frac{\bar{l}-n_c}{\bar{l}-n_c+1}.
\label{rho1}
\end{equation}

\begin{figure}[hbt]
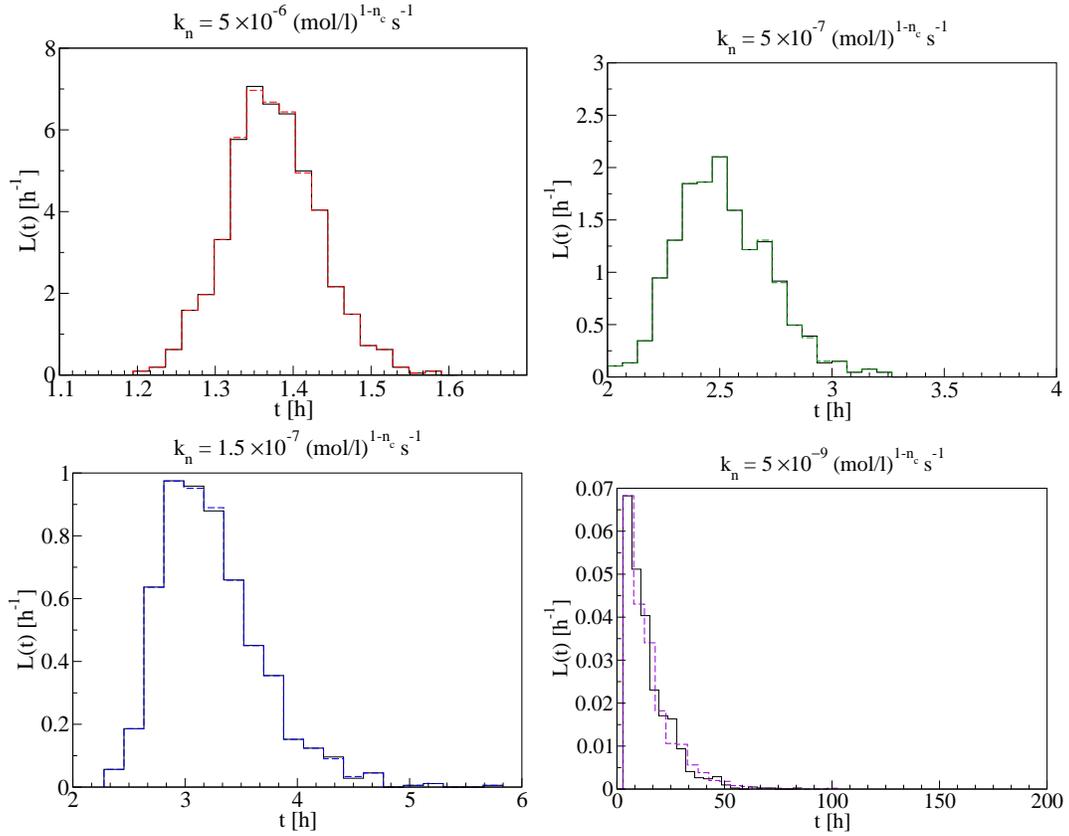

\includegraphics[width=7cm]{fig2_kn_5E-6.eps}
\includegraphics[width=7cm]{fig2_kn_5E-7.eps}\\
\includegraphics[width=7cm]{fig2_kn_1_5E-7.eps}
\includegraphics[width=7cm]{fig2_kn_5E-9.eps}
\caption{\label{fig1_sm} The lag time distribution for several values of $k_n$ for the model including inhomogeneous fragmentation and respecting detailed balance, compared to the original model including nucleation, elongation and (homogeneous) fragmentation only; all the results were obtained by running $1000$ independent kinetic Monte Carlo simulations (using the same random number generator's seeds for both models); see the text below for the values of the other parameters.}
\end{figure}
\noindent Numerical simulations for the stochastic model given by reactions (\ref{cr_nucleation})-(\ref{cr_coalescence}) were performed using the Gillespie algorithm for several nucleation rate constants $k_n$ that yield the same values of $\alpha$ as in Fig. 2 in the main text, using the relationship $\alpha=k_n c_{\mathit{tot}}^{n_c}$; the parameters $V=830$ fl, $M_T=10\%$ of $c_{\mathit{tot}}$, $c_{\mathit{tot}}=100$ $\mu$mol/l, $n_c=2$, $k_+=5\cdot 10^{4}$ l/(mol s) and $k_f=3\cdot 10^{-8}$ $\textrm{s}^{-1}$ were also taken from Fig. 2 in the main text; in addition, the depolymerization rate was set to $k_-=1.1 k_f=3.3\cdot 10^{-8}$ $\textrm{s}^{-1}$; the value for $\rho_1=6.6\cdot 10^{-13}$ mol/l was obtained using (\ref{rho1}) and setting $\bar{l}$ to $1000$; finally, the coalescence and disintegration rates were calculated using (\ref{kc}) and (\ref{kd}) respectively for each of the nucleation rate constants.

Results for the lag time distribution are presented in Fig. \ref{fig1_sm} for the model which includes inhomogeneous fragmentation and respects detailed balance (dashed lines), compared to those for the model analysed in the main text, which assumes homogeneous fragmentation and lacks detailed balance (solid lines). In all four cases the difference between the results of the two models is very small, showing that the effect of inhomogeneous fragmentation and backward processes on the early-time aggregation is negligible. Therefore, the model studied in the main text (Eq. (1)) can safely be used even though it assumes homogeneous fragmentation and neglects detailed balance. 

%------------------------------------------------
% COVARIANCE MATRIX 
%------------------------------------------------

\section{Covariance matrix $\Sigma(t)$}

For completeness, we list here the other matrix elements of $\Sigma(t)$, $\langle x_1(t)^2\rangle$ and $\langle x_1(t)x_2(v)\rangle$, respectively,

\begin{align}
\langle x_1(t)^2\rangle&=\textrm{cosh}(2\tau)\left[\frac{1}{6}\left(\Phi_0+\Psi_0\frac{\lambda}{\mu}\right)+\frac{\alpha n_c}{2\mu}\right]+\textrm{sinh}(2\tau)\sqrt{\frac{\lambda}{\mu}}\left[\frac{\Phi_0+\Psi_0}{3}+\frac{\alpha n_c (n_c-1)}{4\mu}\right]\nonumber\\
&+\frac{\textrm{cosh}\tau}{3\mu}\left(\mu\Phi_0-2\lambda\Psi_0\right)+\frac{\textrm{sinh}\tau}{3}\sqrt{\frac{\lambda}{\mu}}\left(\Psi_0-2\Phi_0\right)-\frac{\lambda}{\mu}\frac{\alpha n_c (n_c-1)t}{2}+\frac{1}{2}\left(\Psi_0\frac{\lambda}{\mu}-\Phi_0-\frac{\alpha n_c}{\mu}\right),\label{x1_2}\\
\langle x_1(t)x_2(t)\rangle&=\sqrt{\frac{\lambda}{\mu}}\left\{\textrm{cosh}(2\tau)\sqrt{\frac{\lambda}{\mu}}\frac{\Phi_0+\Psi_0}{3}+\textrm{sinh}(2\tau)\left[\frac{1}{6}\left(\Phi_0+\Psi_0\frac{\lambda}{\mu}\right)+\frac{\alpha n_c}{2\mu}\right]\right.\nonumber\\
&-\left.\frac{\textrm{cosh}\tau}{3}\sqrt{\frac{\lambda}{\mu}}(\Phi_0+\Psi_0)-\frac{\textrm{sinh}\tau}{3\mu}(\mu\Phi_0+\lambda\Psi_0)\right\}\label{x1x2}.
\end{align}

%------------------------------------------------
% CALCULATION OF THE LAG-TIME DISTRIBUTION L(t)
%------------------------------------------------

\section{Calculation of the lag-time distribution $L(t)$}

to calculate $L(t)$, we first need to find the probability that $m>m_T$ at time $t$, which is given by
\begin{equation}
\textrm{Prob}[m>m_T,t]=\int_{-\infty}^{\infty}dx_1\int_{\frac{m_T\epsilon-\psi(t)}{\sqrt{\epsilon}}}^{\infty}dx_2P(\vec{x},t).
\label{prob_threshold}
\end{equation}
\noindent Here $P(\vec{x},t)$ is a bivariate Gaussian distribution with zero mean and covariance matrix $\Sigma$,
\begin{equation}
P(\vec{x},t)=\frac{1}{2\pi\sqrt{\textrm{det}\Sigma}}e^{-\frac{1}{2}\vec{x}^T \Sigma^{-1}\vec{x}},\quad 
\Sigma=\begin{pmatrix}
\langle x_1(t)^2\rangle & \langle x_1(t)x_2(t)\rangle\\
\langle x_1(t)x_2(t)\rangle & \langle x_2(t)^2\rangle
\end{pmatrix}.
\label{bigauss}
\end{equation}
The argument of the exponential function in (\ref{bigauss}) can be expanded as
\begin{eqnarray}
\vec{x}^T\Sigma^{-1}\vec{x}&=&\frac{\Sigma_{22}x_1^2-2\Sigma_{12}x_1x_2+\Sigma_{11}x_1^2}{\textrm{det}\Sigma}\nonumber\\
&=& \frac{\Sigma_{22}\left(x_1-\frac{\Sigma_{11}}{\Sigma_{22}}x_2\right)^2}{\textrm{det}\Sigma}+\frac{1}{\Sigma_{22}}x_2^2.
\label{bigauss2}
\end{eqnarray}
\noindent The integral over $x_1$ can be easily performed and gives
\begin{equation}
\int_{-\infty}^{\infty}dx_1\exp\left[-\frac{\Sigma_{22}}{2\textrm{det}\Sigma}\left(x_1-\frac{\Sigma_{12}}{\Sigma_{22}}x_2\right)^2\right]=\sqrt{\frac{2\pi\textrm{det}\Sigma}{\Sigma_{22}}},
\end{equation}
\noindent while the integral over $x_2$ can be expressed using the complementary error function
\begin{eqnarray}
\textrm{Prob}[m>m_T,t]&=&\frac{1}{2\pi\sqrt{\textrm{det}\Sigma}}\sqrt{\frac{2\pi\textrm{det}\Sigma}{\Sigma_{22}}} \int_{\frac{m_T\epsilon-\psi(t)}{\sqrt{\epsilon}}}^{\infty}dx_2e^{-\frac{1}{2\Sigma_{22}}x_2^2}\nonumber\\
&=& \frac{1}{\sqrt{\pi}}\int_{\frac{m_t\epsilon-\psi(t)}{\sqrt{2\epsilon\Sigma_{22}}}}dye^{-y^2}\nonumber\\
&=&\frac{1}{2}\textrm{erfc}\left(\frac{m_t\epsilon-\psi(t)}{\sqrt{2\epsilon\Sigma_{22}}}\right).
\end{eqnarray}
\noindent Note that our result for $\textrm{Prob}[m>m_T,t]$ predicts that in general $\textrm{Prob}[m>m_T,\infty]\neq 1$, i.e. not all trajectories will reach the threshold $m_T$. We thus have to count only the trajectories where $m$ will reach $m_T$ \textit{eventually}; this amounts to rescaling $\textrm{Prob}[m>m_T,t]$ with $\textrm{Prob}[m>m_T,\infty]$. 

To find the lag-time distribution $L(t)$, let us look at the probability that $m>m_T$ at time $t+dt$. To find this probability we have to count all events for which $m>m_T$ at time $t$, plus all the events where $m$ has reached $m_T$ in the time interval $[t,t+dt]$. We can thus write
\begin{equation}
\frac{\textrm{Prob}[m>m_T,t+dt]}{\textrm{Prob}[m>m_T,\infty]}=\frac{\textrm{Prob}[m>m_T,t]}{\textrm{Prob}[m>m_T,\infty]}+L(t)dt,
\end{equation}
\noindent from which it follows that 
\begin{equation}
L(t)=\frac{\frac{d}{dt}\textrm{Prob}[m>m_T,t]}{\textrm{Prob}[m>m_T,\infty]}.
\end{equation}
%
%------------------------------------------------
% MOMENTS OF THE LAG-TIME DISTRIBUTION L(t)
%------------------------------------------------

\section{Moments of the lag-time distribution $L(t)$}

In the main text we defined a new variable $r$
\begin{equation}
r=\frac{\psi(t)-m_T\epsilon}{\sqrt{\epsilon\langle x_2(t)^2\rangle}},
\label{r}
\end{equation}
\noindent such that the lag time distribution $L(t)$ becomes a Gaussian in $r$,
\begin{equation}
L(t)dt=\frac{dr/dt}{\sqrt{2\pi}Z}e^{-r(t)^2/2}=\frac{1}{\sqrt{2\pi}Z}e^{-r^2/2}dr.
\end{equation}
\noindent Here $r$ takes values in the range $\langle-\infty,r(\infty)]$, where $r(\infty)$ is given by
\begin{equation}
r(\infty)=\frac{\Psi_0+(\mu/\lambda)\Phi_0}{\sqrt{\epsilon\{[\Psi_0+(\mu/\lambda)\Phi_0]/3+\alpha n_c/\lambda+\sqrt{\mu/\lambda}[(\Psi_0+\Phi_0)/3+\alpha n_c(n_c-1)/(2\mu)]\}}}.
\end{equation}
\noindent To calculate an average of some physical quantity $w(t)$, we can make a change of variable from $t$ to $r$,
\begin{equation}
\langle w(t)\rangle=\int_{0}^{\infty}dt w(t) L(t)=\frac{1}{\sqrt{2\pi}Z}\int_{-\infty}^{r(\infty)}dr w(r) e^{-r^2/2}.
\label{lt_av}
\end{equation}
\noindent To complete the calculation, we have to express $t$ as a function of $r$ so that $w(r)=w(t(r))$. However, the final integral in (\ref{lt_av}) is unlikely to be analytically tractable, and thus we take a different approach. Using the fact that $r(t=T)=0$, we can make a Taylor expansion of $w(t(r))$ around $r=0$,
\begin{equation}
w(t(r))=w(r)=\sum_{k=0}^{\infty}\frac{w^{(k)}(r=0)}{k!}r^k,
\label{t_r_taylor}
\end{equation}
\noindent where $w^{(k)}$ is the $k$-th derivative with respect to $r$. Inserting the Taylor expansion (\ref{t_r_taylor}) in (\ref{lt_av}) we get a formal expression for $\langle w(t)\rangle$ as
\begin{equation}
\langle w(t)\rangle=\sum_{k=0}^{\infty}\frac{w^{(k)}(r=0)}{k!}\langle r^k\rangle
\end{equation}
\noindent where $\langle r^k\rangle$ is given by
\begin{equation}
\langle r^k\rangle=\frac{2^{(k-1)/2}}{\sqrt{2\pi}Z}\left[(-1)^k\Gamma\left(\frac{k+1}{2}\right)+\gamma\left(\frac{k+1}{2},\frac{r(\infty)^2}{2}\right)\right].
\end{equation}
\noindent Here $\Gamma(z)$ and $\gamma(z,x)$ are the gamma and lower incomplete gamma functions, respectively. For example, the first few terms are:
\begin{equation}
\langle r\rangle=-\frac{1}{\sqrt{2\pi}Z}e^{-r(\infty)^2/2},\quad \langle r^2\rangle=1-\frac{r(\infty)}{\sqrt{2\pi}Z}e^{-r(\infty)^2/2}.
\end{equation}
\noindent For higher-order terms we can use the following recursion relation:
\begin{equation}
\langle r^k\rangle=\langle r^{k-2}\rangle-\frac{r(\infty)^{k-1}}{\sqrt{2\pi}Z}e^{-r(\infty)^{2}/2},\quad k\geq 2,
\end{equation}
\noindent which can be solved yielding
\begin{subequations}
\begin{align}
\langle r^{2k}\rangle&=1+\frac{r(\infty)\langle r\rangle (1-r(\infty)^{2k})}{1-r(\infty)^{2}},\quad k\geq 0,\\
\langle r^{2k+1}\rangle&=\frac{\langle r\rangle (1-r(\infty)^{2k+2})}{1-r(\infty)^{2}},\quad k\geq 0.
\end{align}
\end{subequations}
\noindent To complete the calculation of $\langle w(t)\rangle$, we have to calculate the derivatives in (\ref{t_r_taylor}) with respect to $r$, evaluated at $t=T$. These can be found using Fa\`a di Bruno's formula; for example, the first few terms are given by
\begin{eqnarray*}
w^{(1)}(r=0)&=&w^{(1)}(t=T)t^{(1)}(r=0)\\
w^{(2)}(r=0)&=&w^{(2)}(t=T)[t^{(1)}(r=0)]^2+w^{(1)}(t=T)t^{(2)}(r=0)\\
w^{(3)}(r=0)&=&w^{(3)}(t=T)[t^{(1)}(r=0)]^3+3w^{(2)}(t=T)t^{(1)}(r=0)t^{(2)}(r=0)+w^{(1)}(t=T)t^{(3)}(r=0),
\end{eqnarray*}
\noindent where the unknown derivatives $t^{(k)}(r=0)$ can be calculated by setting $w(t)=t$.

By setting $w(t)=t$ and $w(t)=t^2-\langle t\rangle^2$ we get the following expressions for the mean and standard deviation, respectively
\begin{equation}
\langle t\rangle=T-\frac{e^{-r(\infty)^2/2}}{\sqrt{2\pi}Z r^{(1)}(T)}+\left(1-\frac{r(\infty)e^{-r(\infty)^{2}/2}}{\sqrt{2\pi}Z}\right)\frac{r^{(2)}(T)}{2[r^{(1)}(T)]^{3}}+\dots,
\end{equation}
\begin{equation}
\sigma^2=\frac{1}{r^{(1)}(T)}\left(1-\frac{r(\infty)e^{-r(\infty)^{2}/2}}{\sqrt{2\pi}Z}\right)\left[1-\frac{Tr^{(2)}(T)}{r^{(1)}(T)}+\frac{Tr^{(2)}(T)}{[r^{(1)}(T)]^2}\right]+\dots,
\end{equation}
\noindent where $r^{(k)}(T)=\phi^{(k)}(T)/\sqrt{\epsilon\langle x_2(T)\rangle}$ for $k\geq 1$.
%
%--------------------------------------------------------------------------
% REFERENCES
%--------------------------------------------------------------------------

%\begin{thebibliography}{99}
%\bibitem{KnowlesDobson09} T. P. J. Knowles et al., \textit{Science} \textbf{326} 1533-7 (2009)
%\bibitem{MorrisMacPhee13} R. J. Morris,	K. Eden, R. Yarwood, L. Jourdain, R. J. Allen and C. E. MacPhee, \textit{Nat. Commun.} \textbf{4}, 1891 (2013)
%\bibitem{Paturej11} J. Paturej, A. Milchev, V. G. Rostiashvili and T. A. Vilgis, \textit{J. Chem. Phys.} \textbf{134}, 224901 (2011)
%\bibitem{bruno} http://mathworld.wolfram.com/FaadiBrunosFormula.html
%\end{thebibliography}


\begin{thebibliography}{99}
\bibitem{note2} Note that both the deterministic models to which we refer in this paper, and our stochastic model, assume that the system is ``well-mixed'', i.e. spatially homogeneous.
\bibitem{note3} To derive the coarse-grained model from the full one including fragmentation we have to ignore the fact that any fibril created by fragmentation that is smaller than $n_c$ is considered to be unstable and thus dissolves, adding $n_c$ free monomers to the solution.
\bibitem{note4} The variable $r(t)$ has the properties that $r(0)=-\infty$ and $r(T)=0$; also, $r(\infty)$ is finite and can be calculated by noting that for large times $\psi(t)\approx A\exp(\sqrt{\mu\lambda}t)$ and $\langle x_2^2(t)\rangle\approx C\exp(2\sqrt{\mu\lambda}t)$ yielding $r(\infty)=A/\sqrt{\epsilon C}$.
\bibitem{note5} In Ref. \cite{Knowles11}, $k_+$ and $k_f$ were calculated from the speed and width of the aggregation front in the droplets; we also assumed, as in previous work that $n_c=2$ and $\phi(0)=\psi(0)=0$ \cite{KnowlesDobson09,MorrisMacPhee13}.
\end{thebibliography}
\end{document}